\newcommand{\N}{I\!\!N}

\def\tf{t_s(r)}
\documentclass[oneside,draft,a4paper,10pt]{amsart}

\usepackage{amsmath}               
\usepackage{amscd}
\usepackage{amsthm}                
\usepackage{times}

\numberwithin{equation}{section}

\title[Naked singularities in dust collapse as an existence problem for O.D.E.]%
{Naked singularities in dust collapse as an existence problem for O.D.E. at a singular point}
\author[R.\ Giamb\`o]{Roberto Giamb\`o}
\address{Dipartimento di Matematica e Fisica\hfill\break\indent
Universit\`a di Camerino, Italy}
\email{roberto.giambo@unicam.it}
\urladdr{http://www2.unicam.it/\~{}giambo}
\author[G.\ Magli]{Giulio Magli}
\address{Dipartimento di Matematica,\hfill\break\indent Politecnico di Milano, Italy}
\email{magli@mate.polimi.it}

\begin{document}

\date{September 2001 (accepted for publication on Diff. Geom.
Appl.)}


\swapnumbers

\theoremstyle{plain}\newtheorem{teo}{Theorem}[section]
\theoremstyle{plain}\newtheorem{prop}[teo]{Proposition}
\theoremstyle{plain}\newtheorem{lem}[teo]{Lemma}
\theoremstyle{plain}\newtheorem{cor}[teo]{Corollary}
\theoremstyle{definition}\newtheorem{defin}[teo]{Definition}
\theoremstyle{remark}\newtheorem{rem}[teo]{Remark}
\theoremstyle{plain} \newtheorem{assum}[teo]{Assumption}
\theoremstyle{definition}\newtheorem{example}[teo]{Example}

\begin{abstract}

The final state of the gravitational collapse of a marginally bound dust cloud
is formulated in terms of an existence problem for the non-linear differential
equation governing radial null geodesics near the singular point.
Rigorous results are proved,
covering the complete spectrum of the possible initial data.

\end{abstract}

\maketitle

\begin{section}{Introduction}\label{sec:intro}

It is well known
that stable, non singular states of superdense matter can exist
only if the mass of the final object is less than a physical
limit, namely the Chandrasekar limit (about $1.4 M_{\odot}$) in
the case of white dwarfs or the neutron star limit (of the order
of $3M_{\odot}$) in the case of neutron stars. For collapsing
objects which are unable to radiate away a sufficient amount of
mass to fall below such limits, no final stable state is available
and therefore singularities are formed.

A famous conjecture, first formulated by Roger Penrose \cite{Penrose69}
and known as the {\it Cosmic Censorship} conjecture
states that a blackhole is always formed in complete gravitational collapse of reasonable matter fields.
However, if stated without any further mathematical assumption,
the conjecture is false, since several examples of {\it naked singularities},
i.e. solutions of the Einstein field equations describing
singularities not hidden behind an absolute event horizon, are known.
It is, therefore, of primary importance to understand the mathematical structure
of such singularities, with the final aim of reformulating
the conjecture as a theorem and hopefully prove it.

Examples of focussing naked singularities in gravitational collapse
firstly arose from numerical investigations by Eardley  \cite{E7}
and Eardley and Smarr \cite{ES},
while the first to perform a
formal investigation was
Christodoulou \cite{C1}. In his paper,  Christodoulou used a fixed point
technique to show that the equation of radial null geodesics
for a collapsing dust ball starting form rest and having a parabolic density profile
has a solution meeting the singularity in the past, the latter being thus ``visible"
to nearby observers. Since then, a technique has been developed which
makes use of L'Hopital theorem to identify existence of solutions with finite tangent
near the singularity (``root equation" approach, see e.g. \cite{JD}). In particular,
all the possible endstates of the gravitational collapse of spherically symmetric dust
have been obtained in this way \cite{Jos}, as well as the final states of
gravitating systems of rotating particles known as Einstein clusters \cite{JM}.
The root equation technique, however, proves useful only if the
exact explicit solution of the Einstein field equations is known for the case at hand.
As a consequence, we are still very far from a complete understanding of
the Censorship problem even in the simple case of spherical symmetry, since
very few exact solutions are known. In addition, the root equation
approach is essentially related to the existence of solutions of a
specific kind, that is not {\it a--priori} guaranteed.

In this paper, we give a o.d.e. approach to the nature of the singularities
in marginally bound dust collapse. Using classical techniques
we make rigorous, by explicit construction,
the results obtained previously with the root equation technique.

\end{section} 

\begin{section}{Collapsing dust clouds in General Relativity}\label{sec:dust}

A collapsing sphere of dust in General Relativity is described
by the famous solution which brings the names of Lemaitre, Tolman and Bondi
(see e.g. \cite{Kramer}).
We concentrate here only the case in which the cloud is marginally bound
(the velocity is zero at space infinity). Using comoving coordinates,
the metric is
$$
ds^2=-dt^2+(R')^2
dr^2+R^2(d\theta^2+\sin^2\theta d\phi^2)
$$
(we denote by a prime and a dot the partial derivatives with respect to $r$ and $t$).
The function $R=R(r,t)$ satisfies the Kepler-like
equation of motion $\dot R^2=2F(r)/R$ and is therefore
given by
\begin{equation}\label{eq:R}
R(r,t)=r(1-k(r)t)^{\frac 23}
\end{equation}
where $k(r)=(3/2)\sqrt{2F(r)/r^3}$.
In the above formulae, $F(r)$ is the initial distribution of mass
of the cloud (and thus is a positive function).
The energy density is given by
$$
\epsilon (r,t) =\frac{F'}{4\pi R^2R'}
$$
at $t=0$ one has $
\epsilon (r,0) =\frac{F'}{4\pi r^2}
$ and therefore regularity of the Cauchy data at $r=0$ implies $F \approx r^3$ as $r$
tends to zero. We assume (as usual) the function $F(r)$
to be Taylor-expandable near $r=0$ (all our results actually hold true
also if $F$ is only of class $C^3$).
Therefore we put
$$
F(r)=F_0r^3+F_nr^{n+3}+\Gamma(r)
$$
where $\Gamma(r)$ is infinitesimal of order greater than or equal
to $n+4$. The physical requirement that the density has to be positive and
decreasing outwards
further imply that $F_0$ is positive and $F_n$ is negative.
It follows easily that
\begin{equation}\label{eq:kappa}
k(r)=1-a\,r^n+\gamma(r),
\end{equation}
where $\gamma(r)$ is infinitesimal of order greater than or equal
to $n+1$, $a$ is some positive constant and, without loss of generality,
we have put $k(0)=1$.

The energy density becomes
singular whenever $R$ or $R'$ vanish during the evolution.
Thus, singularities can be of
two different kinds: {\it shell crossing}, at which
$R'$ vanishes  while $R$ is non-zero, and {\it shell focusing}
at which $R$ vanishes. The shell crossing
singularities have been frequently considered as ''weak'' although
no proof of extensibility is as yet available in the literature.
In any case, in most physically interesting situations such
singularities do not occur, so that we shall concentrate attention
here only on the shell focussing case.

The locus of the zeroes of the function $R(r,t)$ defines the {\it
singularity curve} $\tf $ by the relation $R(r,\tf )=0$. Due to formula
\eqref{eq:R}, we have $\tf =1/k(r)$.
Physically, $\tf$ is that comoving time at which the shell of
matter labeled by $r$ becomes singular. The singularity
forming at $r=0,t=t_s (0)$ is called central and, in dust clouds,
is the unique singularity that can be naked.
To see this, we recall that a singularity cannot be naked if
it occurs after the formation of the apparent horizon.
The apparent horizon  ($t_h(r)$, say) is the boundary of the region
of trapped surfaces and is defined by
the equation $R(r,t_h(r)=2F(r)$, that is

\begin{equation}\label{eq:horizon}
t_h(r)=\tf -\frac{8}{27} k(r)^2 r^3
\end{equation}
so that $\tf > t_h(r)$ for any $r>0$.

To analyze the causal structure of the
central singularity, observe that, if
the singularity is visible, at least one outgoing null geodesic
must exist, that meets the singularity in the past. Such a
geodesic will be a solution of
\begin{equation}\label{f}
\frac{dt(r)}{dr}=\varphi (r,t)
\end{equation}
where
\begin{equation}\label{eq:phi}
\varphi(r,t):=\sqrt{
-\frac{g_{rr}}{g_{00} } }=
\frac{1-k(r)\,t-\frac 23 r\,k'(r)\,t}{(1-k\,t)^{1/3}}
\end{equation}
with initial datum $t(0)=t_s(0)=1$.
For a problem of this kind, in which the initial point is singular (the function $\varphi$ is not
defined at $(0,t(0))$) no general results of existence/non existence are known.
As a consequence, in the literature, an approach has been developed \cite{JD} which makes
use of l'Hopital theorem to identify the possible values
of the tangent of the geodesic curve at the singularity.
What turns out is the following:
\begin{itemize}\item
For $n=1$ or $n=2$ the singularity is naked;
\item
For $n=3$ the singularity is naked
if $a\ge a_c$ where
\begin{equation}\label{ac}
a_c=\frac{2(26+15\sqrt{3})}{27}.
\end{equation}
Therefore, $a_c$ is a ``critical parameter":
at $a=a_c$ a ``phase transition'' occurs and
the endstate of collapse turns from a naked singularity
to a blackhole.
\item
If $n>3$ the singularity is covered.
\end{itemize}

This approach, however, strictly depends on the form of the
solution of \eqref{f}, that must be of the form $1+x r^\alpha$
with $x$ constant. Anyway, the root equation can be as well
recovered following our approach for proving nakedness,
where we will look for solutions
of the form $1+x(r) r^\alpha$, and impose a continuity condition
on the unknown function $x(r)$.

\end{section} 

\begin{section}{Non-existence}\label{sec:non-ex}

We begin by stating the non-existence result.
The argument covers the case $n\ge 4$ and
gives a partial answer in the case $n=3$,
(the remaining part is given in Section \ref{sec:n=3}).

\begin{teo}\label{teo:non-ex1}
If $n\ge 4$ the singularity is covered.
\end{teo}

To prove the above statement we need the following:

\begin{lem}\label{lem:subsol}
There exists $r_*>0$ such that the apparent horizon $t_h(r)$ is a
subsolution of \eqref{f}
for $r\in(0,r_*)$.
\end{lem}
\begin{proof}
Recall that $t_h(r)=\frac1{k(r)} - \frac{8}{27} k(r)^2 r^3$, and
$k(r)\cong 1- a r^n$, where in last relation $a>0$ and
infinitesimal of order greater than $n$ has been dropped. We must
show that $t_h(r)\le\varphi(r,t_h(r)),\,\forall r\in (0,r_*)$ with
$r_*$ sufficiently small. One gets
\begin{equation}\label{eq:dtdr}
\frac{\text dt_h}{\text dr}=
-\frac{k'}{k^2} -\frac{8}{27}(2 k k' r^3 + 3 k^2 r^2),
\end{equation}
and
\begin{equation}\label{eq:phith}
\varphi(r,t_h(r))
=-\frac{k'}{k^2}+\frac{4}{9} k^2 r^2 + \frac{8}{27}k k' r^3,
\end{equation}
where the relation
\[
1-k t_h=\frac{8}{27}k^3 r^3
\]
has been used. It follows
\begin{equation}\label{eq:t'-phi}
\frac{\text dt_h}{\text dr}-\varphi(r,t_h(r))=-\frac{4}{3} k r^2
\left(\frac{2}{3} k' r + k\right),
\end{equation}
that is negative for $r$ sufficiently small and positive.
\end{proof}

{\sl Proof of Theorem \ref{teo:non-ex1}.}
Let $t_\rho(r)$ the solution of $t'(r)=\varphi(r,t(r))$ such that
$t_\rho(0)=1$.
By contradiction we suppose
the existence of $r_1>0$ such that $t_\rho(r_1)<t_h(r_1)$ and
$t_\rho(r)\le t_h(r),\,\forall r\in [0,r_1]$. We can suppose
$r_1<r_*$, where $r_*$ comes from Lemma \ref{lem:subsol}.
Since $t_\rho(0)=t_h(0)$, one has
\begin{multline}\label{eq:contra1}
0<t_h(r_1)-t_\rho(r_1)=\left(t_h(r_1)-t_h(0)\right)-\left((t_\rho(r_1)-t_\rho(0))\right)=\\
\left(t'_h(\xi)-t'_\rho(\xi)\right)\,r_1=(t'_h(\xi)-\varphi(\xi,t_\rho(\xi)))\,r_1,
\end{multline}
where $\xi\in(0,r_1)$. Using Lemma \ref{lem:subsol} it is
$t'_h(\xi)\le \varphi(\xi,t_h(\xi))$, and hence
\begin{equation}\label{eq:contra2}
t'_h(\xi)-\varphi(\xi,t_\rho(\xi))\le\frac{\partial\varphi}{\partial
t}(\xi,\theta)\,\,(t_h(\xi)-t_\rho(\xi)).
\end{equation}
Combining \ref{eq:contra1} and \eqref{eq:contra2} one gets a
contradiction if $\frac{\partial\varphi}{\partial
t}(\xi,\theta)\le 0$. Now:
\begin{equation}\label{eq:dphidt}
\frac{\partial\varphi}{\partial t}(r,t)=
-\frac{2}{3}
\frac{k (1- k t)+ r k' \left(1- \frac{2}{3} k t\right)}{(1- k
t)^{4/3}}.
\end{equation}
Hence $\frac{\partial\varphi}{\partial t}(r,0)<0$ for
small values of $r$. Showing
that $\frac{\partial\varphi}{\partial
t}(r, t_h(r))\le 0$
we have that $\frac{\partial\varphi}{\partial t}
(\xi,\theta)\le 0$, since we can observe that the numerator in last
term of \eqref{eq:dphidt} is linear in $t$ and the denominator has
a fixed sign.
Using \eqref{eq:dphidt} we have
\begin{equation}\label{eq:dphidt(th)}
\frac{\partial\varphi}{\partial
t}(r, t_h(r))=-\frac{2}{3} \frac{\frac{8}{27} k^4 r^3+k'r-\frac
{2}{3}k k' r t}{\left(\frac{2}{3}k r\right)^4}=
-\left(\frac 23\right)^{-3}
\left(\frac {8}{27} r^{-1}-\frac 13 n\,a\, r^{n-4}\right),
\end{equation}
where infinitesimal of order greater than $n$ has been dropped out
in last quantity,
so that the sign of the right hand side in \eqref{eq:dphidt(th)}
depends on $n$, and is strictly negative
if $n\ge 4$.
\qed\bigskip

The above argument provides only a sufficient
condition for nakedness. Indeed, it does not
exhaust all cases for the singularity to be covered (see Section
\ref{sec:n=3}).
\end{section}

\begin{section}{Existence}\label{sec:ex}

In this section we establish rigorously existence of naked
singularities in the cases $n=1,2$.

\begin{teo}\label{thm:ex12}
If $n=1,2$ there exists a geodesics of the form
\begin{equation}\label{eq:sol}
t_\rho(r)=1+x(r) r^\alpha,\qquad r\in [0,r^*],
\end{equation}
where $\alpha=1+{\frac 23}n$ and $x(r)$ is a differentiable
function in $[0,r^*]$ such that $x(0)>0$.
\end{teo}

\begin{proof}
We will show the existence of a function $x(r)\in H^{1,p}[0,r^*]$ with
$p>1$ and
$r^*>0$ sufficiently small, and of a parameter $\alpha\ge 1$
such that $x(0)>0$ and $t=1+x\,r^\alpha$
solves equation $t'(r)=\varphi(r,t(r))$, where $\varphi(r,t)$ is
given by \eqref{eq:phi}:
\begin{equation}\label{eq:ODE}
t'(r)=\frac{1-k(r)\,t(r)-\frac 23
r\,k'(r)\,t(r)}{(1-k\,t(r))^{1/3}},\qquad k(r)=1-a\,r^n+\gamma(r).
\end{equation}
We recall that $\gamma(r)$ is infinitesimal of order greater than $n$
as $r\to 0^+$.
Substituting in \eqref{eq:ODE} the expressions for $t(r)$ and
$k(r)$, and using that $n\ge 1$ one gets
\begin{equation}\label{eq:ODEx}
r\,x'= r^{\frac 23 n+1-\alpha}\left[
\frac
{a + \frac 23 an + r b(r)+ x c(r) r^\alpha- d(r) x r^{\alpha-n}}
{\left(a + a x r^\alpha - x r^{\alpha-n}-\delta(r)(1+x r^\alpha)\right)^{1/3}}
\right] - \alpha\,x,
\end{equation}
where $b(r), c(r)$ are continuous functions differentiable in $r=0$,
$d(r)=1+\gamma(r)$, $\delta(r)$ is an infinitesimal differentiable
function of order greater than or equal to $1$ for $r\to 0^+$, and $\alpha$ is a positive
parameter to be determined below.
We search for $x$ such that $r\,x'$ is infinitesimal for $r\to 0^*$. Then the
right hand side of \eqref{eq:ODEx} must be infinitesimal for $r\to
0^+$. Since the quantity in square brackets is bounded for
$n=1,2$, it must be ${\frac 23 n+1-\alpha}\ge0$. But if the strict
inequality held, the limit of the left hand side would be $-\alpha
x(0)$ which by hypothesis is strictly negative. So the only
possible situation is
$\alpha=1+\frac 23 n$
from which one gets
\begin{equation}\label{eq:root12}
x(0)=a^{\frac 23}
\end{equation}
using the infinitesimal behaviour of the left hand side of
\eqref{eq:ODEx}.

Having chosen the values of $\alpha$ and the initial condition
$x(0)$, one has actually to show the existence of a solution of
\begin{multline}\label{eq:ODE3}
r\,x'=\left[a\,\left(1+\frac 23 n\right)+r\,b(r)+x c(r) r^{1+\frac 23 n}-\right.\\
-\left.d(r) x r^{1-\frac 13 n} -\left(1+\frac 23 n\right) x\, G(x,r)\right] G^{-1}(x,r),\qquad
x(0)=a^{\frac 23},
\end{multline}
where
\begin{multline}\label{eq:G}
G(x,r)=a^{1/3}\left[1+\left(-\frac{x r^{1-\frac 13 n}}{a}+x\,r^{1+\frac 23
n}+\frac\delta a+\frac\delta a x r^{1+\frac 23
n}\right)\right]=\\
=a^{1/3}\left[1+\frac 13\left(-\frac{x r^{1-\frac 13 n}}{a}+x\,r^{1+\frac 23
n}+\frac\delta a+\frac\delta a x r^{1+\frac 23
n}\right)+ A(r) r^{2\left({1-\frac 13 n}\right)}\right].
\end{multline}
We observe that last relation has been written using Taylor
expansion of the quantity in round bracket in the first row of
\eqref{eq:G}, and in view of this the continuous function $A(r)$
has been introduced.
Using \eqref{eq:G} in \eqref{eq:ODE3}, and collecting terms with
the same power of $x$ the differential equation becomes
\begin{multline}\label{eq:ODE4}
r\, x'=G(x,r)^{-1}\left[a\left(1+\frac 23 n\right)-a^{1/3}\left(1+\frac 23
n\right)x+\right. \\
\left.+r\,b(r) + \left(e(r)\,x+f(r) x^2\right)r^{1-\frac13
n}\right],\qquad x(0)=a^{2/3},
\end{multline}
where $e(r),\,f(r)$ are continuous functions, differentiable in
$r=0$. It is a straightforward calculation that $e(0)=-1,\,f(0)>0$.

With the positions
\[
y=x-a^{2/3},\qquad \beta=\alpha-n=1-\frac 13 n,
\]
one recovers a differential equation of the form
\begin{equation}\label{eq:ODEfin}
r y'= A(r,y) y + B(r,y) r^\beta,\qquad y(0)=0,
\end{equation}
where $A(r,y)$ and $B(r,y)$ are continuous functions such that
$A(0,0)<0$ and $B(0,0)>0$.

Let us now define four constants that bound $A$ and $B$ in a small
neighborhood $\mathcal U=[0,r^*]\times[-\epsilon,\epsilon]$
of $(r,y)=(0,0)$:
\[
A_0\le A(r,y)\le A_1<0,\quad 0<B_0\le B(r,y)\le B_1,\qquad
(r,y)\in\mathcal U.
\]
Let us also define the two positive functions
\begin{equation}\label{eq:z0z1}
z_0(r)=\frac{B_0}{\beta- A_0} r^\beta,\quad z_1(r)=\frac{B_1}
{\beta- A_1} r^\beta,\qquad r\in [0,r^*]
\end{equation}
respectively solutions of the Cauchy problems
\begin{equation}\label{eq:ODEz}
\begin{cases}
z'=\frac 1r\,A_0 z+ B_0 r^{\beta-1},\\ z(0)=0
\end{cases}
\qquad
\begin{cases}
z'=\frac 1r\,A_1 z+ B_1 r^{\beta-1},\\ z(0)=0
\end{cases}
\end{equation}
It is readily observed that $z_0(r)<z_1(r)\,\forall r\in [0,r^*]$.
Hence, $\forall n\in\N$, let $y_n$ denote the solution of the ODE in
\eqref{eq:ODEfin} with the initial condition $y(\frac 1n)=y_{0n}$
such that
\[
z_0(\frac 1n)\le y_{0n} \le z_1(\frac 1n).
\]
From comparison theorems in ODE one gets
\begin{equation}\label{eq:comp}
0< z_0(r)\le y_n(r)\le z_1(r),\qquad r\in [1/n,r^*],
\end{equation}
and then extending $y_n$ to $[0,r^*]$ setting $y_n=y_{0n}$ in
$[0,\frac 1n]$ we have that $|y_n|$ are equibounded by $K\, r^\beta$
with $K$ constant. Moreover, using the ODE in \eqref{eq:ODEfin} $|y'_n|$
are equibounded by $K\, r^{\beta-1}$ which is
$L^p,\,p>1$. So, up to
subsequences, $y_n$ converges uniformly to a function $y$ in
$H^{1,p}$, which is easily shown to be a differentiable solution of \eqref{eq:ODEfin}
using the ODE in \eqref{eq:ODEfin} and Lebesgue theorem.
\end{proof}

\end{section} 

\begin{section}{The critical case}\label{sec:n=3}
The analysis so far shows existence of naked
singularities if $n=1,2$, and non--existence if $n>3$. When $n=3$ a
partial answer is contained in Theorem \ref{teo:non-ex1}.
Indeed, the key point in the proof is the study of the sign of
\eqref{eq:dphidt(th)} for small values of $r$. In the case $k=1-a
r^3$, omitting infinitesimal of order greater that $3$, direct
substitution in \eqref{eq:dphidt(th)} yields
\begin{equation}\label{eq:dphidt(th)-n=3}
\frac{\partial\varphi}{\partial
t}(r, t_h(r))=-\left(\frac{2}{3}\right)^{-3} \left(\frac{8}{27}-a\right) r^{-1},
\end{equation}
Then we
must impose the condition $\frac{8}{27}-a\ge 0$ in order to recover
the same situation as in Theorem \ref{teo:non-ex1}. In other
words, we have shown the following
\begin{prop}\label{prop:nonex-n=3}
If $n=3$ and $a\le\frac{8}{27}$ the singularity is covered.
\end{prop}

Sufficient conditions to ensure existence of naked singularity can
now be  given, with a repetition of the argument used in Theorem
\ref{thm:ex12}. In this case one can show the existence of a
solution of the kind $t(r)=1+x(r) r^3$, with $x(r)\in H^{1,p}[0,r^*],\,p>1$
and $x(0)>0$.
Since $\alpha-n=0$ we must be careful in treating the
infinitesimal terms in the differential equation \eqref{eq:ODEx}, which
now takes the form
\begin{equation}\label{eq:ODEx-n=3}
r\,x'=\left[\frac{3a - d(r) x+ r b(r)+ x c(r) r^3}{
\left(a - x + a x r^3 -\delta(r)(1+x
r^3)\right)^{1/3}}\right]-3 x,
\end{equation}
where $b(r),c(r)$ and $d(r)$ have the same meaning as in
\eqref{eq:ODEx}. In order to ensure the infinitesimal behaviour of
the right hand side of \eqref{eq:ODEx-n=3}, we must then require
\[
\frac{3a-x(0)}{(a-x(0))^{1/3}}-3x(0) =0.
\]
Since we want $x(0)>0$, this implies that $a$ must be such that
the algebraic equation
\begin{equation}\label{eq:root}
27 x^3 (a-x) - (3a - x)^3=0.
\end{equation}
has real positive roots. It is a simple exercise to check that
this is true only if $a\le a_0 $
or $a\ge a_c$, where $a_0=(2/27)(26+15\sqrt{3})^{-1}$ while
$a_c$ is defined in \eqref{ac}.
The first case however must be
excluded since  the solution would not live below the apparent
horizon $t_h(r)$. Indeed, we know from Proposition \ref{prop:nonex-n=3}
that the singularity is covered if $a<\frac{8}{27}$.
We must instead accept the second interval, and the same
arguments of Theorem \ref{thm:ex12} can be used with some
slight modifications here, in order to ensure the following
\begin{prop}\label{prop:ex-n=3}
If $n=3$ and
$a\ge a_c$ the singularity is naked.
\end{prop}

What remains to be analyzed is
whether naked
singularities may exist for $a\in(\frac{8}{27},a_c)$.
Actually, such solutions represent blackholes, since we can
show that the sufficient condition of Proposition \ref{prop:ex-n=3} is
also necessary in this case.
\begin{prop}\label{prop:nonex-gap}
If $n=3$ and the singularity is naked, then $a\ge a_c$.
\end{prop}
\begin{proof}
Let $t_\rho(r)$ be a solution of the differential equation
$t'_\rho(r)=\varphi(r,t_\rho(r))$. We can write it
in the form $t_\rho(r)=1+x(r) r^3$,
although in this case we don't know the behaviour of $x(r)$ near
the origin $r=0$. We just know $x$ continuous, $x(r) r^3\to 0$ as $r\to 0^+$ and,
since the singularity is naked,
$t_\rho(r)\le t_h(r)=1+(a-\frac{8}{27})r^3+o(r^3)$. Last fact implies
\begin{equation}\label{eq:upper}
x(r)\le a-\frac{8}{27}+\eta,
\end{equation}
for $r>0$ sufficiently small, where $\eta\ll 1$ is a constant.
Then $x(r)$ is bounded from above in a right neighborhood of $r$.
But it is also bounded from below. Indeed, $t'_\rho(r)> 0$
since $\varphi(r,t_\rho(r))> 0$ for $r>0$ small,
and then $x(r) r^3$ is increasing. Thus $x(r) r^3$
must approach $0$ from above as $r\to 0^+$, and then $x(r)$ must
be positive, and therefore bounded.

Now let us write \eqref{eq:ODEx-n=3} as
\begin{equation}\label{eq:ODEroot}
r\,x'=\frac{3a -x + f(r)}{(a-x+g(r,x))^{1/3}}
-3x
\end{equation}
where $f(r)\to 0$ and, since $x r^3\to 0$,
also $g(r,y)\to 0$ per $r\to 0^+$.

Moreover, let us define
\[Q(a,x)=\frac{3a -x}{(a-x)^{1/3}} -3x.
\]
Recall that \eqref{eq:upper} ensures that $(a-x)^{1/3}>0$ for $r$
small.
Then \eqref{eq:ODEroot}
may be written as
\begin{equation}\label{eq:Q}
Q(a,x)=r x'-h(r,x),
\end{equation}
with $h(r,x)\to 0$ as $r\to 0^+$.

At this point we don't know whether $\lim_{r\to 0} x(r)$ exists or
not. If it does, then $r x'(r)$ tends to $0$. Indeed, from
\eqref{eq:Q} we get that $r x'(r)$ tends to $Q(a,x(0))$; if this
quantity was not null, then $x'(r)$ would behave like $\frac 1r$ in
a right neighborhood of $0$, and then $x(r)$ would not be bounded,
behaving like $\log r$. Then $Q(a,x(0))=0$, which means that $a$ is such
that $x(0)$ is a positive root of the equation \eqref{eq:root}.

If $\lim_{r\to 0} x(r)$ does not exist, since $x$ is bounded there
must exists a sequence $(r_n,x_n=x(r_n))$ with
$r_n\to 0$ and $x'(r_n)=0$ as $n\to\infty$. This shows that $\{x_n\}$
is such that $Q(a,x_n)\to 0$. Up to subsequences, $\{x_n\}$
converges to a positive root of \eqref{eq:root}.

\end{proof}

The fact that we were obliged to divide the analysis on the critical cases
into two intervals of values of $a$ is the mathematical
reflection of interesting physical phenomenon \cite{co}.
In fact, we are using formation of the apparent horizon to obtain non existence.
Absence of apparent horizon
is only a necessary condition for nakedness, and in fact there
is a interval of values  of $a$ for which the singularity is not visible, but the slope
of the apparent horizon does allow for a geodesic to come out.

\end{section} 
\vskip .5 truecm

{\bf Acknowledgment.} The authors wish to thank Fabio Giannoni
for useful discussions and suggestions.
\medskip

{\bf Note added.} After completion of this work we became aware of
related independent paper by Mena and Nolan \cite{Nolan}.

\end{document}